\documentclass[aps,prl,reprint,floatfix,preprintnumbers,nofootinbib,a4paper,superscriptaddress,10pt]{revtex4-2}

\usepackage{amssymb,amsmath,verbatim,mathtools,needspace,enumitem,etoolbox,graphicx,microtype}
\usepackage{multirow}
\usepackage[dvipsnames, usenames]{xcolor}
\usepackage{bigints}

\definecolor{linkcolor}{rgb}{0.0,0.4,0.4}
\definecolor{citecolor}{rgb}{.7,.3,.5}
\usepackage[colorlinks=true, linkcolor=linkcolor, citecolor=citecolor, pdfusetitle]{hyperref}
\usepackage[
  left=0.7in,
  right=0.7in,
  top=0.8in,
  bottom=1in
]{geometry}
\usepackage[all]{hypcap}
\usepackage[T1]{fontenc}
\usepackage[utf8]{inputenc}
\usepackage{graphicx,amssymb,amsmath,amsthm,amsfonts,epsfig,times,natbib}
\usepackage[normalem]{ulem}
\usepackage{multirow}
\usepackage{import}
\usepackage[titletoc,title]{appendix}

\usepackage{graphicx}				
\usepackage{placeins}
\usepackage{subcaption}
\usepackage{ragged2e} 
\usepackage{array}
\DeclareCaptionFormat{myformat}{\justifying#1#2#3}
\usepackage{verbatim}
\usepackage{rotating}
\usepackage{comment}
\usepackage{txfonts}

\usepackage{epstopdf}
\usepackage{tensor}
\usepackage{amsbsy}
\usepackage{bm,url}
\usepackage{float}
\usepackage{dcolumn}
\usepackage{latexsym}
\usepackage{rotating}
\usepackage{enumerate}
\def\be{\begin{equation}}
\def\ee{\end{equation}}
\def\bea{\begin{eqnarray}}
\def\eea{\end{eqnarray}}

\newcommand{\ba}{\begin{align}}
\newcommand{\ea}{\end{align}}

\newcommand{\deco}{\texttt{PhenomDECO}}
\newcommand{\bbh}{ \texttt{PhenomD}}

\newcommand{\AEI}{Max Planck Institute for Gravitational Physics (Albert Einstein Institute),
Callinstrasse 38, D-30167 Hannover, Germany}
\newcommand{\Leibniz}{Leibniz University Hannover, 30167 Hannover, Germany}
\newcommand{\Portsmouth}{Institute of Cosmology and Gravitation, University of Portsmouth, Portsmouth, PO1 3FX, UK}
\newcommand{\Cardiff}{Gravity Exploration Institute, Cardiff University, Cardiff, United Kingdom}

\begin{document}
\title{ Establishing Compactness as a Population Observable in Gravitational-Wave Astronomy }
\author{Shrobana Ghosh}
\email{shrobana.ghosh@aei.mpg.de}
\affiliation{\AEI}
\affiliation{\Leibniz}
\author{Charlie Hoy}
\email{charlie.hoy@port.ac.uk}
\affiliation{\Portsmouth}

\author{Mark Hannam} 
\affiliation{\Cardiff}
\author{Frank Ohme}
\affiliation{\AEI}
\affiliation{\Leibniz}

\begin{abstract}
Classically, black holes (BHs) are the most compact objects predicted in nature with C=0.5 in the Schwarzschild limit; C is defined as the mass-to-radius ratio in geometric units. In this work we perform a novel measurement on the nature of putative BH mergers in the gravitational wave (GW) data by directly probing the binary’s closest approach through an effective compactness parameter. We confidently show all such high-significance signals in GWTC-3 are consistent with the BH hypothesis for the first time. Our hierarchical analysis yields $C_{\rm eff} = 0.5^{+0.3}_{-0.1}$, and we further limit the merger rate of low-compactness exotic binaries to $< 0.7\,{\rm Gpc}^{-3}\,{\rm yr}^{-1}$. This work establishes compactness as a key observable in GW astronomy.

\end{abstract}

\maketitle

\paragraph{\textbf{Introduction} --} Compact objects are among the most interesting predictions of general relativity (GR): sufficiently massive stars are expected to undergo gravitational collapse and leave behind ultra-dense remnants~\cite{PhysRev.56.455}, with BHs being the simplest classical vacuum solution. There are also a number of alternative compact-object models that depart from the classical BH picture, either by replacing the event horizon with horizonless structure, as in boson stars~\cite{Kaup1968,RuffiniBonazzola1969}, traversable wormholes~\cite{MorrisThorne1988}, and fuzzball~\cite{Mathur2005} constructions, or by resolving the central singularity through a regular interior geometry, as in gravastars~\cite{MazurMottola2004}, with a de Sitter core and a thin shell of matter. Such exotic compact objects (ECOs) have been proposed in a variety of theoretical settings, including quantum gravity, modified gravity, and dark matter (DM) scenarios. Therefore, establishing whether any such objects exist in nature would inform not only the astrophysics of compact remnants formed through stellar collapse, but also point to possible new physics from the early universe or beyond the Standard Model.

A useful way to investigate the nature of compact objects is in terms of compactness, which characterizes the ratio of mass $m$ to size  $r$, i.e., $C = m/r$, which is dimensionless in geometric units. Indirect constraints on compactness have long been pursued through electromagnetic observations, including X-ray binaries~\cite{Narayan:2005ie}, stellar orbits around Sgr A*~\cite{Gillessen:2008qv,GRAVITY:2020gka} and, more recently, horizon-scale imaging of supermassive BHs~\cite{EventHorizonTelescope:2019ggy,EventHorizonTelescope:2022wkp}. Depending on the system and observable, these probes access regions ranging from a few to thousands of gravitational radii. In the supermassive regime, stellar dynamics and BH imaging probe the near-horizon spacetime relatively directly, whereas for stellar-mass X-ray binaries the inference is more indirect, relying on accretion disk modeling and arguments based on quiescent luminosity or the absence of detectable surface emission~\cite{McClintock:2003gx,McClintock:2004ji}, all of which are subject to astrophysical and modeling uncertainties. By contrast, gravitational waves (GWs) probe the strong-field dynamics of compact binaries in the near-merger regime, where source compactness can directly affect the observed signal morphology. Compactness-based arguments have already played an important role in GW interpretation, most notably for GW150914~\cite{LIGOScientific:2016wyt}, where the inferred mass scale and merger frequency implied a closest approach comparable to the Schwarzschild scale. However, compactness has not yet been established as a generic observable that can be inferred and reported systematically across GW catalogs~\cite{GWTC5}. Given the broad and poorly constrained space of theoretical alternatives to binary black holes (BBHs), a theory-agnostic approach is well suited: rather than targeting a specific ECO model, one can ask whether the observed signals are compatible with BBH-like merger morphology. Measuring compactness at the population level is also important in GW inference, where single-event posteriors can be affected by statistical fluctuations. It allows us to test whether non-BBH-like support emerges collectively across events, providing observation driven guidance for future ECO models.

\begin{figure*}[t!]
	\includegraphics[width=0.95\textwidth]{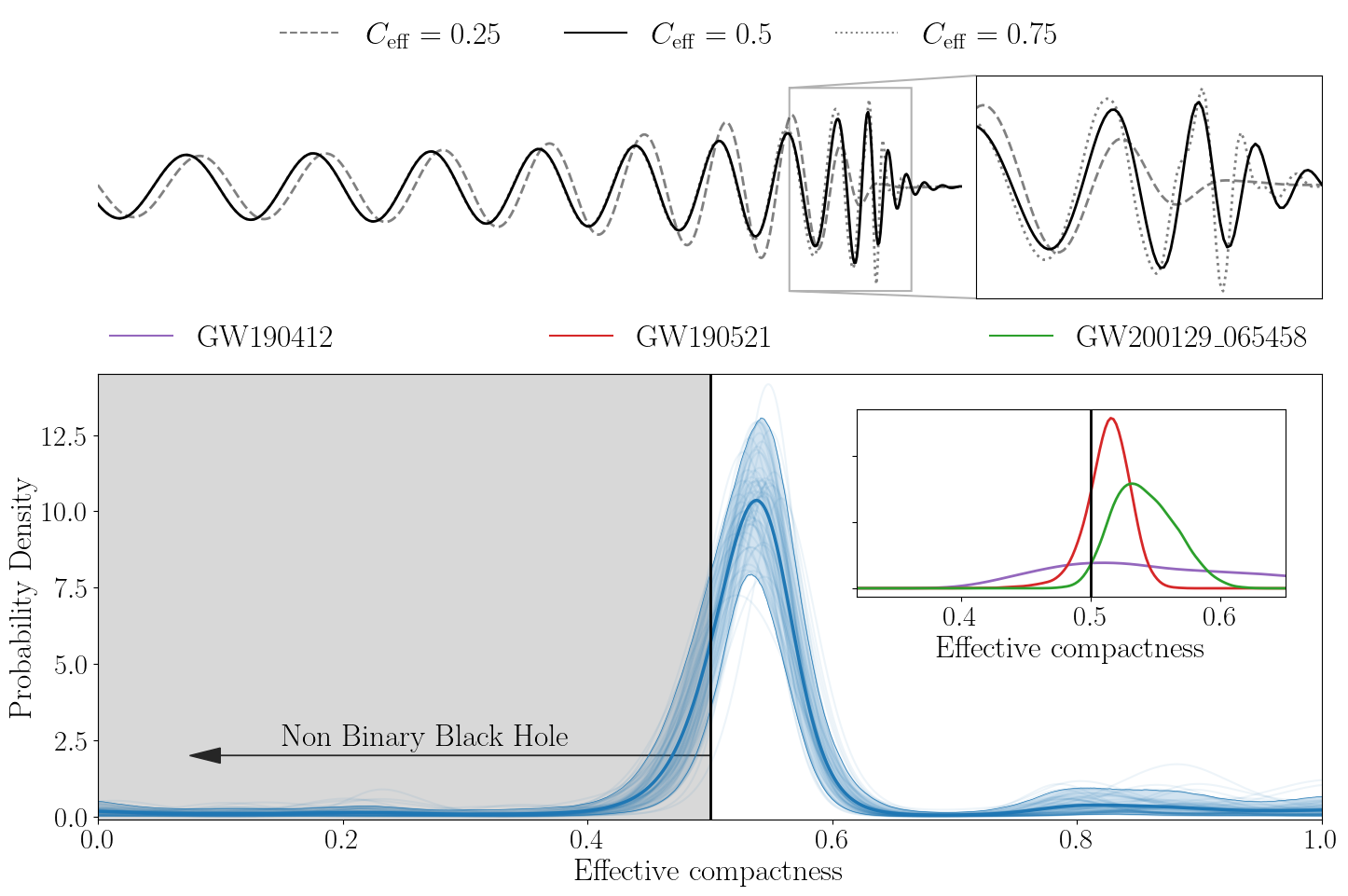}
	\caption{The first population-level measurement for effective compactness from GW observations. The thick blue line shows the median of the posterior predictive distribution for the astrophysical compactness of black hole binaries in GWTC-3, with the shaded region showing the 90\% credible interval. The light blue traces show individual draws from the posterior. The solid black vertical line shows the expectation for a binary black hole, and the shaded grey region shows the compactness for exotic compact objects. The top right inset highlights the inferred compactness for several events that the GW community have singled out as being \emph{exceptional signals} -- GW190412~\cite{LIGOScientific:2020stg}, GW190521~\cite{LIGOScientific:2020iuh,LIGOScientific:2020ufj} and, GW200129\_065458~\cite{Abbott_2023}. Above the figure we show how the effective compactness changes the observed GW signal for a fiducial binary black hole produced with~\deco. We consider nonspinning BHs with component masses $20\, M_{\odot}$ and $10\, M_{\odot}$ and show only the plus polarization for clarity. }
\label{fig:compactness_population}
\end{figure*}

As a first step in this direction, Ref.~\cite{Ghosh:2025wex} introduced~\deco, an extension to a standard BBH waveform model to infer an effective compactness from GW data. The effective compactness parameter characterizes how close the binary components are at merger, or more precisely when the waveform departs from the BBH expectation. Directly defining this separation in coordinates is subtle in GR, since coordinate separations depend on gauge choices and are difficult to interpret physically in the strong-field regime near merger; moreover, such a coordinate separation is not itself an observable in GW data analysis. Instead,~\deco~exploits the characteristic near-merger structure of the frequency-domain amplitude in compact-binary signals. By tracking this morphology through the observed GW frequency,~\deco~maps departures from BBH-like behaviour to an effective contact separation for non-BH objects. This takes closest-approach reasoning beyond heuristic estimates and poses it as an inference problem: effective compactness and source masses are inferred simultaneously, which is essential because ECO signals analysed with BBH templates can be absorbed into biased mass parameters~\cite{Ghosh:2025wex}. We direct readers to a companion article~\cite{companion_article} for a more detailed discussion on the effective compactness parameter.

Here, we apply~{\deco} to infer the effective compactness for a set of compact-binary events representative of the BBH population observed to date. At the level of individual events, compactness inference is limited by noise fluctuations, waveform systematics, and data-quality issues. A hierarchical population analysis helps to convert a set of individual constraints into a statement about the underlying merger population: whether it is concentrated near the BBH value or shows evidence for a sub-population with non-BBH-like compactness. Although LIGO-Virgo-Kagra (LVK) analyses have performed several tests relevant to the nature of compact objects~\cite{LIGOScientific:2020tif,LIGOScientific:2021sio}, but these tests are not designed to provide a single source-property measurement that can be uniformly reported and combined across events. In this Letter, we test whether the compact binary mergers in GWTC-3.0~\cite{Abbott_2023} are collectively consistent with a population of BBH-like compactness. This approach is naturally most sensitive to the dominant population, while any rare outliers require separate treatment through more targeted analyses, as discussed in the companion article~\cite{companion_article}, along with more detailed assessment of the impact of non-gaussianities in detector noise and waveform systematics. 
Nonetheless, we do not find strong evidence for ECOs in any single GW observation. The resulting population distribution is of interest not only astrophysically, but also methodologically: it provides an empirically informed prior on compactness for future GW inference.

\paragraph{\textbf{~\deco~as a compactness probe}--} The signal from the merger of exotic compact objects may deviate from that of coalescing BHs through several finite-sized effects, such as tidal deformability, spin-induced quadrupole moments, and modified absorption/heating effects associated with the presence or absence of horizons~\cite{Chia:2022rwc,Sennett:2017etc,Krishnendu:2017shb,Johnson-McDaniel_2020,Datta:2019epe}. 
All of these effects can lead to tiny deviations in the phase of the signal as has been shown in numerical simulations of boson star binaries~\cite{Palenzuela:2006wp,Palenzuela:2007dm,Dietrich:2018bvi,Clough:2018exo,Evstafyeva:2024qvp}. Such deviations can be picked up by current GW detectors~\cite{LIGOScientific:2014pky,aLIGO:2020wna,Tse:2019wcy,VIRGO:2014yos,Virgo:2019juy,Virgo:2022ysc,Somiya:2011np,PhysRevD.88.043007,KAGRA:2020tym}, provided that the signal is loud enough. In fact, several of the reported GW events have been investigated for such deviations~\cite{Krishnendu:2019tjp,PhysRevLett.126.081101,PhysRevLett.125.261105,Narikawa:2021pak,Chia:2023tle,krishnendu2025testingnaturecompactobjects,Datta:2020gem,Krishnendu:2025rud}. For objects with significantly different compactness than BHs, this dephasing may also be accompanied by contact established between the objects, instead of a plunge, at a frequency lower than the merger frequency for BBHs.
~\deco~ attempts to capture this early end of inspiral by modifying the morphology of the signal amplitude in an underlying BBH model, \bbh~\cite{Khan:2015jqa}. 
~\deco~does not measure the component compactnesses. Instead, it constrains an effective compactness of the binary at contact, $C_{\rm eff}$, where $C_{\rm eff} = 0.5$ recovers the BBH case. 
As was previously demonstrated~\cite{Ghosh:2025wex},~\deco~can directly infer the posterior probability distribution on compactness from real events without significantly biasing other source parameters.


\paragraph{\textbf{Dataset} --}
We define our event set using the same astrophysical significance threshold adopted in the GWTC-3 population analysis, selecting putative BBH mergers with false-alarm rate $<0.25 \rm{yr^{-1}}$; this yields 69 signals detected by the LIGO Hanford, LIGO Livingston and Virgo detectors. We exclude BNS or NSBH mergers because current observations do not generally resolve their merger/post-merger signal, which is crucial for constraining effective compactness with~\deco. Adopting the same event definition as the GWTC-3 population analysis allows the compactness inference to be interpreted directly in the context of established population results. We focus on publicly released data from GWTC-3~\cite{gwosc_gwtc} because it already captures the broad population trends seen in BBH mergers to date (cf. see Fig.~4 in Ref.~\cite{LIGOScientific:2025pvj}).

\paragraph{\textbf{Population Inference} --} We provide the first population-level constraint on an effective compactness of assumed BBH mergers by performing a hierarchical Bayesian analysis that combines single-event-level source properties. To obtain estimates for the compactness of each GW source, we perform Bayesian inference with~\deco~over a 12-dimensional parameter space: the individual component masses, spin magnitudes, effective compactness, inclination angle, luminosity distance, right ascension, declination, phase, polarization and merger time. Owing to the large parameter space making the posterior distribution analytically intractable, we use the {\textsc{Dynesty}} nested sampling algorithm~\cite{Speagle:2019ivv} to draw samples from the unknown posterior via the {\textsc{Bilby}} library~\cite{Ashton:2018jfp, Romero-Shaw:2020owr}, as has been done in all LVK analyses since the third GW catalog~\cite{Abbott_2023}. We employ the same priors, power spectral densities, calibration envelopes and settings as was done in the original analyses performed by the LVK where the BBH hypothesis was assumed. The only exceptions are that we employ a uniform prior distribution for the effective compactness of the binary, ranging from 0.1 to 0.99, and that we use stricter settings to ensure sampler convergence. Additional details are provided in the Appendix.

Once we obtain individual source properties for all events, we employ a \emph{weakly modelled} approach to estimate the population level compactness measurement through the {\textsc{gwpopulation}} library~\cite{2019PhRvD.100d3030T,Talbot2025}: we use a piecewise polynomial model -- a cubic spline -- with 14 nodes to ensure minimal assumptions about the underlying astrophysical population. We use 4 ghost nodes lying outside the $0 - 1$ region to control edge effects, 2 above and 2 below, and 10 nodes within the astrophysical region. We use agnostic priors for the amplitude of each node by assuming a unit Gaussian distribution, as was done in e.g. Ref.~\cite{Golomb:2022bon}. We also checked the robustness of our population result by varying the  number of nodes, and consistent results were obtained. Additional details are provided in the Appendix.

\paragraph{\textbf{Results} --} From Fig.~\ref{fig:compactness_population}, we see that the compact binary mergers observed by the LVK in GWTC-3.0 are consistent with a compactness of 0.5 within the 90\% credible interval (CI); from the median of the posterior predictive distribution, we infer a compactness of $C=0.54^{+0.28}_{-0.12}$ with the median measured to an accuracy of $C = 0.54^{+0.01}_{-0.01}$(all reported errors are quoted as symmetric 90\% CI unless otherwise stated). We find no significant evidence that compact objects observed in GWTC-3 are exotic within this parametrisation. Given current detector sensitivities, our agnostic population analysis estimates that $80^{+15}_{-26}\%$ of binaries have compactness $\geq 0.5$. This increases to $96^{+2}_{-4}\%$ for compactness $\geq 0.4$; this more conservative boundary was chosen to account for small shifts in the single-event distributions due to prior choices, noise or missing physics in the model, see  Ref~\cite{companion_article} for details.

While not a significant departure, we observe a tightly constrained compactness measurement that peaks above 0.5. Under the assumption that all events feature the same type of compact object, we may simply multiply the likelihoods from the single-event posterior distributions. In that case, we similarly observe a shift to $C = 0.52^{+0.01}_{-0.01}$. This is due to most single-event distributions showing broad support for $C > 0.5$, with a subset showing either a bimodal distribution with a secondary peak at $\sim 0.8$ or a railing distribution at $C \sim 1$. Although $C_{\rm eff} < 0.5$ can be interpreted as non-BH compactness, we interpret all larger values as being consistent with the BH hypothesis. Indeed, we find a strong degeneracy between all signals with $C_{\rm eff} > 0.5$.

\begin{figure}[t!]
	\includegraphics[width=0.4\textwidth]{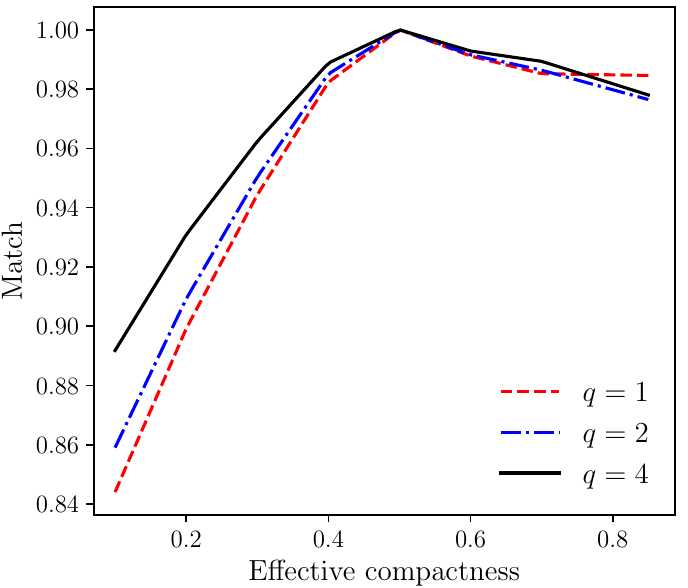}
	\caption{Noise-weighted matches between BBH signals and \deco\ waveforms, computed using the Advanced LIGO sensitivity for  $M=40 M_\odot$ and $q\in[1,2,4]$ and~\deco~show that lower-compactness waveforms are more readily distinguishable from the C=0.5 case than higher-compactness waveforms.}
\label{fig:matchC}
\end{figure}

In Fig.~\ref{fig:matchC}, we show that the noise-weighted matches~\cite{Owen:1995tm} between~\deco~waveforms generated with $C_{\rm eff}>0.5$ are much closer to the BBH case than those generated with $C_{\rm eff}<0.5$. A match of unity indicates identical waveforms, while a match near zero indicates highly dissimilar waveforms. Here we compute the match over the frequency range of $20 -1024$ Hz using the Advanced LIGO sensitivity curve $S_n(f)$~\cite{adLIGOnoise} using the standard match function from the \texttt{PyCBC} library. This is further emphasised in the top panel of Fig.~\ref{fig:compactness_population}, which shows marginal difference between the GW signals produced with compactness $C = 0.5$ and $C = 0.75$, but significant differences for $C < 0.5$.  We therefore do not interpret support at $C>0.5$ as evidence for ordinary horizonless compact objects, and attribute it to noise fluctuations and/or waveform systematics. For example, the support for $C_{\rm eff}>0.5$ seen in the compactness posteriors for GW200129\_065458, shown in the inset of Fig.~\ref{fig:compactness_population}, may be due to lack of spin-precession in the baseline model that~\deco~uses, since this event shows significant evidence for misaligned spins~\cite{Hannam:2021pit,Hoy:2024wkc}. For a more stringent astrophysical interpretation, future analyses could therefore consider using a population-informed prior on effective compactness~\cite{Mould:2026nle}, restricting it to the upper $90\%$ CI of the hierarchical inference result or, more conservatively, to $C_{\rm eff}\leq0.5$.

Finally, for about a dozen events, we find that our analysis fails to fit the astrophysical signal. We infer an effective compactness posterior sharply peaked at $C\simeq0.14$--$0.17$ with a poorly localised geocentric coalescence time and, in most cases, a high total mass. This is indicative of the analysis fitting non-gaussian noise features in the data. Through performing single-detector analyses with different frequency content, we attribute the origin of this behaviour to the LIGO Livingston data below $\sim 50\, \mathrm{Hz}$. Repeating the analysis when neglecting LIGO Livingston data below $50\,{\rm Hz}$ removes spurious signals and recovers BBH-like effective compactness for all affected events. This is explained in more detail in a companion paper~\cite{companion_article}. 


Focusing on GW190521, an event that has been considered for different exotic scenarios~\cite{PhysRevLett.126.081101,PhysRevLett.125.261105,Aurrekoetxea:2023vtp}, we find the compactness measurement to strongly peak at $C \sim 0.5$, revealing that the source is likely a BBH (when assuming a compact-binary origin). We note that previous work has argued that GW190521 could have been produced from the head-on collision of Proca stars~\cite{PhysRevLett.126.081101}. However, Proca stars are expected to be less compact than BHs~\cite{Herdeiro:2016tmi}.

\begin{figure}[t!]
	\includegraphics[width=0.48\textwidth]{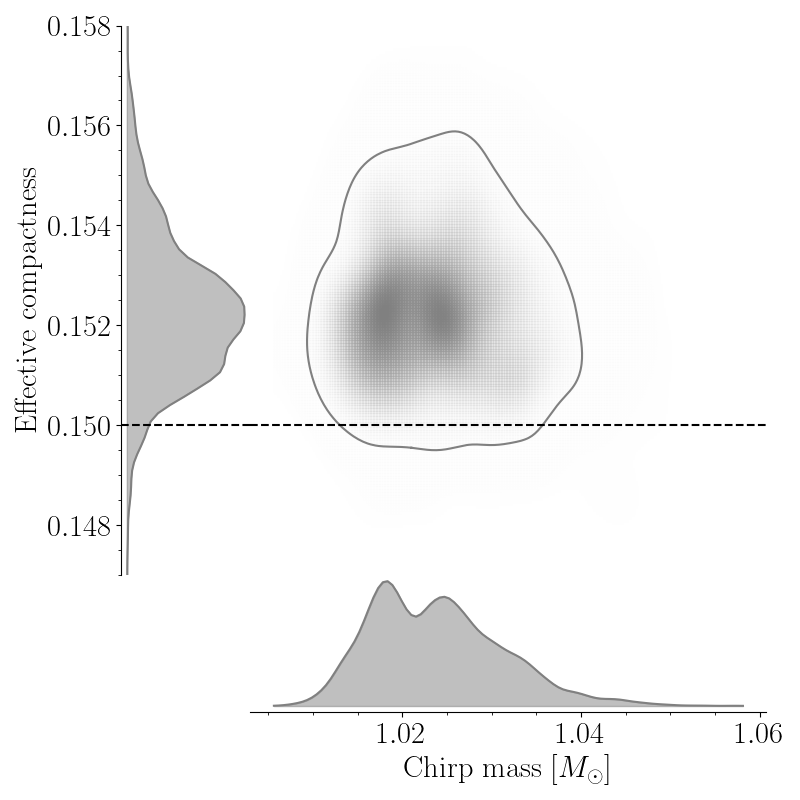}
	\caption{Effective compactness and chirp mass estimates obtained with~\deco~for a numerical-relativity BNS signal injected into noise of NEMO, a hypothetical detector sensitive to BNS mergers. We consider the BNS-0094 simulation with true chirp mass $1.16\, M_{\odot}$ and an estimated compactness $\sim 0.15$ (shown by the black dashed line). The contour in the central panel shows the 90\% credible interval.}
\label{fig:bns_compactness}
\end{figure}

Next, we confirm that~\deco~is not preferentially biased toward recovering BBH-like effective compactness. To illustrate this, we show in Fig.~\ref{fig:bns_compactness} the result of analysing a binary neutron star (BNS) signal with~\deco. Specifically, we inject the BNS-0094 numerical-relativity waveform from the CoRe database of BNS waveforms~\cite{Dietrich:2018phi} into a synthetic noise realization for the proposed detector NEMO~\cite{Ackley:2020atn}, which is designed to be sensitive to the BNS merger and post-merger signal, and analyse the data over the frequency range $410\text{--}4096\,\mathrm{Hz}$. GW signals in the LIGO and Virgo detectors are typically analysed from $20\,\mathrm{Hz}$, but $410\,\mathrm{Hz}$ was used due to limited inspiral of the numerical-relativity waveform. We find that~\deco~confidently disfavors the BBH hypothesis with an inferred effective compactness well below $0.5$.

To assess whether the this result is reasonable for the BNS waveform, we compare it with a simple contact-scale estimate. We compute the ratio of the total mass to a near-contact separation between the stellar centers to estimate a geometric measure of the binary's compactness. Approximating this separation as the sum of the stellar radii,
and using the individual stellar compactnesses, $C_{i} = m_{i}/r_{i}$ where i $\in \{A,B\}$ for each star, we estimate,
\begin{equation}
    C_{\rm geometric} = \frac{m_A + m_B}{m_A/C_A + m_B/C_B}.
    \label{eq:ceff}
\end{equation}
Both neutron stars in the NR injection are modeled with the MS1b equation of state~\cite{Mueller:1996pm,Read:2008iy}. For the specified component masses this fixes the tidal Love numbers, ($k_2^A = 0.087$ and $k_2^B = 0.16$) and dimensionless tidal deformabilities, ($\Lambda_2^A = 182.8$ and $\Lambda_2^B = 9280$). Following the standard relation
\begin{equation}
\Lambda_2 = \frac{2 k_2}{3C^5},
\end{equation}
we have $C_A\sim0.2$ and $C_B\sim0.1$. Plugging these in Eq.~\eqref{eq:ceff} and noting that $m_A = 1.94 M_\odot$ and $m_B=0.94 M_\odot$, we get a $C_{\rm{geometric}}\sim0.15$. While the inferred compactness appears consistent with this estimate, we observe a slight bias in the chirp mass and more pronounced biases in the mass ratio (with the posterior peaking at $q=4$) and effective spin when using~\deco. These biases are likely driven by the absence of tidal effects in the model, as also noted in~\cite{Dudi:2018jzn}.

\paragraph{\textbf{ECO Rates} --} We found no viable ECO candidates in our analysis. 
Based on this null result, we can derive a rough upper estimate of the ECO merger rate. Assuming a Poisson distribution for the number of detected ECO binaries in an observing period~\cite{Mandel:2018mve}, which accounts for the random occurrence times of mergers and for the fact that only a fraction
of the underlying population is detectable, the probability of observing $k$ such events can be written as
\begin{equation}
    p(k|N_{\rm ECO})
    =
    \frac{N_{\rm ECO}^{k} e^{-N_{\rm ECO}}}{k!},
    \label{eq:poisson_eco}
\end{equation}
where $N_{\rm ECO}$ denotes the expected number of detected ECO binaries in a search. From the present analysis, we set $k=0$, and Eq.~\eqref{eq:poisson_eco} gives
\begin{equation} 
p(0|N_{\rm ECO}) = e^{-N_{\rm ECO}}. 
\end{equation}
Requiring $p(0|N_{\rm ECO})>0.1$ gives $N_{\rm ECO}<2.3$ at 90\% confidence. Thus, any ECO population model that predicts more than 2.3 detectable events with effective compactness significantly lower than BBHs in the O3 BBH-candidate sample, under a BBH-like mass, spin, and redshift distribution, is disfavoured at 90\% confidence. Translating this bound into an astrophysical ECO merger rate requires the ECO sensitive spacetime volume, $\langle VT\rangle_{\rm ECO}$,~\cite{LIGOScientific:2016kwr} which must be computed with ECO-specific waveform injections under the assumed population model. As an illustrative example if we assumed $\langle VT\rangle_{\rm ECO}^{\rm O3}\simeq\langle VT\rangle_{\rm BBH}^{\rm O3}$ this would correspond to an upper limit of the order $2.3/69\simeq3\%$ of the BBH merger rate i.e., $0.72\, \rm{Gpc}^{-3}\,\rm{yr}^{-1}$~\cite{LIGOScientific:2025pvj}.
\\
\\

\paragraph{\textbf{Conclusion} --} The compact binaries observed to date are identified primarily through searches using BBH waveform models. Consequently, the main marker of their likeness to a BBH merger morphology has been whether a BBH waveform in GR provides a sufficiently good fit to the data. While such goodness-of-fit measures are essential, they do not by themselves define a source property that can be inferred, reported, and combined across a catalog. A poor BBH fit may arise from noise, calibration uncertainty, waveform systematics, precession, eccentricity, or non-BBH physics, and therefore does not directly quantify the degree to which the merger morphology is BBH-like. 

In this work we have shown that by introducing a phenomenological deformation to the BBH merger morphology,~\deco~provides such a complementary observable via an effective compactness parameter. This is distinct from previous parameterized tests that introduce deviations, for example in the phase evolution, but do not track a physically motivated source property associated with the merger morphology that can be reported as a catalog-level BBH-likeness measure. Applied to 69 putative BBH mergers from GWTC$-3.0$, our hierarchical analysis gives the first population-level measurement of an effective merger compactness in GW observations. The inferred population is sharply concentrated near the BBH expectation, $C\simeq0.5$, with no evidence for a significant population of lower-compactness driven earlier mergers.

We have also shown, using a numerical-relativity BNS waveform, that~\deco~is not intrinsically biased toward the BBH value. When the signal contains genuinely non-BBH-like merger information, the model can recover an effective compactness significantly below $C=0.5$. At present,~\deco~should be regarded as complementary to dedicated BNS and NSBH waveform models: those models include tidal physics, while $\deco$ provides a phenomenological compactness observable tied to the merger morphology. More accurate compactness measurements of exotic binaries will require waveform models that include both tidal effects and, where relevant, post-merger phenomenology, with consistent parametrization of compactness through the full signal. 

Finally, the absence of confident low compactness events already places a simple constraint on the number of detectable ECO binaries in the analyzed sample. Assuming Poisson counting statistics, any ECO population model that predicts more than 2.3 detectable events with low effective compactness in the O3 BBH-candidate sample, under BBH-like mass, spin, and redshift distributions, is disfavoured at 90\% confidence. Translating this bound into an astrophysical ECO merger rate requires the ECO sensitive spacetime volume, \(\langle VT\rangle_{\rm ECO}\), computed with ECO-specific waveform injections. From the set of observations we have analysed we estimate an upper limit on low-compactness ECO mergers of $0.7\,{\rm Gpc}^{-3}\,{\rm yr}^{-1}$. Improved rate estimates could be made by considering the full set of GW observations to date, improved ECO signal modelling, and a detailed calculation of the sensitive spacetime volume for ECO signals. The catalog-level effective compactness measurement introduced here therefore provides both a new empirical baseline for the BBH-like merger population and a target for future ECO population models to predict, test, and constrain their merger rates. 

\vspace{2em}

We thank N. V. Krishnendu for comments during the LIGO--Virgo--KAGRA internal review, and Laura Nuttall and K. G. Arun for comments on the results of this project. SG and FO acknowledge support from the Max Planck Society and CH thanks the University of Portsmouth for support through the Dennis Sciama Fellowship. MH was supported by Science and Technology Facilities Council (STFC) grants ST/V005618/1, ST/Y004272/1 and UKRI2489. We are grateful for computational resources provided by the LIGO Laboratory, supported by the National Science Foundation Grants PHY-0757058 and PHY-0823459, and the SCIAMA high performance computing cluster supported by the Institute of Cosmology and Gravitation (ICG) and the University of Portsmouth.

This research has made use of data or software obtained from the Gravitational Wave Open Science Center (gwosc.org), a service of the LIGO Scientific Collaboration, the Virgo Collaboration, and KAGRA. This material is based upon work supported by NSF's LIGO Laboratory which is a major facility fully funded by the National Science Foundation, as well as the Science and Technology Facilities Council (STFC) of the United Kingdom, the Max-Planck-Society (MPS), and the State of Niedersachsen/Germany for support of the construction of Advanced LIGO and construction and operation of the GEO600 detector. Additional support for Advanced LIGO was provided by the Australian Research Council. Virgo is funded, through the European Gravitational Observatory (EGO), by the French Centre National de Recherche Scientifique (CNRS), the Italian Istituto Nazionale di Fisica Nucleare (INFN) and the Dutch Nikhef, with contributions by institutions from Belgium, Germany, Greece, Hungary, Ireland, Japan, Monaco, Poland, Portugal, Spain. KAGRA is supported by Ministry of Education, Culture, Sports, Science and Technology (MEXT), Japan Society for the Promotion of Science (JSPS) in Japan; National Research Foundation (NRF) and Ministry of Science and ICT (MSIT) in Korea; Academia Sinica (AS) and National Science and Technology Council (NSTC) in Taiwan. This material is based upon work supported by NSF's LIGO Laboratory which is a major facility fully funded by the National Science Foundation. For the purpose of open access, the author(s) has applied a Creative Commons Attribution (CC BY) licence to any Author Accepted Manuscript version arising.

\section{Appendix}

\paragraph{\textbf{Gravitational-wave Bayesian Inference} --} Assuming a GW signal is of astrophysical origin, a probability distribution for the source properties $\boldsymbol{\lambda} = \{\lambda_{1} + \lambda_{2} + ...\}$ can be estimated through single-event Bayesian inference. Here, the model-dependent \emph{posterior} probability distribution can be inferred through,

\begin{equation}
    P(\boldsymbol{\lambda} | d) = \frac{\mathcal{L}(d | \boldsymbol{\lambda}, \mathfrak{m})\, \Pi(\boldsymbol{\lambda} | \mathfrak{m})}{\mathcal{Z}},
\end{equation}
where $d$ is the GW data -- comprising signal $h$ and noise $n$ -- $\mathfrak{m}$ is a parameterised GW model, $\mathcal{L}(d | \boldsymbol{\lambda}, \mathfrak{m})$ is the probability of obtaining the data given the source properties $\boldsymbol{\lambda}$ and model $\mathfrak{m}$, otherwise known as the likelihood, $\Pi(\boldsymbol{\lambda} | \mathfrak{m})$ is the probability of obtaining the source properties $\boldsymbol{\lambda}$ given the model, otherwise known as the prior, and $\mathcal{Z} = \int{\mathcal{L}(d | \boldsymbol{\lambda}, \mathfrak{m})\, \Pi(\boldsymbol{\lambda} | \mathfrak{m})} d\boldsymbol{\lambda}$, otherwise known as the evidence.

Although the likelihood is well-known in GW astronomy, see e.g. Ref.~\cite{Veitch:2014wba}, it is often not computationally possible to analytically calculate $P(\boldsymbol{\lambda} | d)$. This is because the  evidence requires computing a high dimensional integral over a wide parameter space. As such, stochastic sampling methods are often used to draw samples from the unknown posterior probability distribution. Commonly used techniques include Markov-Chain Monte-Carlo (MCMC)~\cite{metropolis1949monte} and nested sampling~\cite{Skilling2004,Skilling:2006gxv}. For the case of nested sampling, a set of live points are randomly drawn from the prior and, through an iterative process, these live points converge to the high likelihood region(s). In more detail, during each iteration the live point with the lowest likelihood is stored, and replaced with another drawn from within the {\emph{likelihood-constrained prior}} -- a subset of the prior which has likelihood greater than the point it is replacing. This process continues until a stopping criterion is reached.

Numerous packages are available to perform GW Bayesian inference~\cite{Veitch:2014wba,Ashton:2018jfp,Romero-Shaw:2020owr,Biwer:2018osg,Lange:2018pyp,Roulet:2022kot,Tiwari:2023mzf}, with {\textsc{bilby}}~\cite{Ashton:2018jfp, Romero-Shaw:2020owr} being the flagship code used by the LVK, and used to analyse all GW signals since the third GW catalog~\cite{Abbott_2023}. {\textsc{bilby}} interfaces with many public MCMC~\cite{Foreman-Mackey:2012any,Ashton:2021anp} and nested samplers~\cite{Speagle:2019ivv,Williams:2021qyt,del2022cpnest}. In the main paper, we analysed all significant GW signals in GWTC-3 with the {\textsc{Dynesty}}~\cite{Speagle:2019ivv} nested sampler via {\textsc{bilby}}.

Typically, GW Bayesian Inference is performed via nested sampling with 1000 live points~\cite{LIGOScientific:2025yae}. In the main paper, we used 2000 live points to improve sampler convergence; more live points has been shown to reduce implementation-specific eﬀects, as well as errors from the stochastic nature of nested sampling~\cite{Higson:2018cqj}. All other settings matched the original analyses performed by the LVK where the simplified BBH hypothesis was assumed. We also assumed the same priors as those used by the LVK, except we projected them into the non-precessing space, i.e. we marginalized over the in-plane spin directions, since~\deco~does not incorporate precession effects. 

\paragraph{\textbf{Hierarchical Bayesian Inference} --} Once a catalog of GW signals has been detected, $\{D\}$, and single-event Bayesian inference has been performed on each, hierarchical Bayesian inference can be performed to infer the properties of the underlying population $\boldsymbol{\Lambda} = \{\Lambda_{1}, \Lambda_{2}, ...\}$. A posterior probability distribution for the population properties can be inferred through,

\begin{equation}
    p(\boldsymbol{\Lambda} | \{D\}) = \frac{\mathcal{L}(\{D\} | \boldsymbol{\Lambda})\, \Pi(\boldsymbol{\Lambda})}{\mathfrak{Z}},
\end{equation}
where $\mathcal{L}(\{D\} | \boldsymbol{\Lambda})$ is the probability of the measurements given the hyper parameters $\boldsymbol{\Lambda}$, otherwise known as the hyper likelihood, $\Pi(\boldsymbol{\Lambda})$ is the probability of the hyper parameters $\boldsymbol{\Lambda}$, otherwise known as the hyper prior and $\mathfrak{Z} = \int \mathcal{L}(\{D\} | \boldsymbol{\Lambda})\, \Pi(\boldsymbol{\Lambda})\, d\boldsymbol{\Lambda}$ otherwise known as the hyper evidence. Assuming all candidates are of astrophysical origin, the hyper likelihood is~\cite{Mandel:2018mve}

\begin{equation}
    \mathcal{L}(\{D\} | \boldsymbol{\Lambda}) = \prod_{i=1}^{N_{\mathrm{obs}}} \frac{\int d{\boldsymbol{\lambda}}_{i}\,\mathcal{L}(D_{i}|{\boldsymbol{\lambda}}_{i})\,\Pi({\boldsymbol{\lambda}}_{i}|{\boldsymbol{\Lambda}})}{\int{d\boldsymbol{\lambda}}_{i}\,p_{\mathrm{det}}({\boldsymbol{\lambda}}_{i})\,\Pi({\boldsymbol{\lambda}}_{i}|{\boldsymbol{\Lambda}})},
\end{equation}
where $D_{i}$ is the $i$th event in the catalog, and $p_{\mathrm{det}}$ is the detection probability of a GW with parameters ${\boldsymbol{\lambda}}_{i}$, and $\Pi(\boldsymbol{\lambda}_{i} | \boldsymbol{\Lambda})$ is the prior probability of drawing  a GW with parameters  ${\boldsymbol{\lambda}}_{i}$ from a population with hyperparameters $\boldsymbol{\Lambda}$.

The detection probability accounts for the fact that existing GW detectors are more sensitive to certain GW signals than others. For example, we are more likely to observe high-mass, aligned spin systems since their SNR is significantly larger than a low-mass, negatively aligned spin system at a fixed distance. This means that our detectors can observe high-mass binary systems significantly further away, and hence, the associated search volume is much larger. The detection probability is typically calculated by simulating millions of GW signals from a known population, injecting them into real GW strain data, and searching for them using known techniques~\cite{Farr:2019rap,Tiwari:2017ndi}.

Similar to the single-event analogue, in order to perform Hierarchical Bayesian inference, the {\textsc{gwpopulation}}~\cite{2019PhRvD.100d3030T,Talbot2025} and {\textsc{bilby}}~\cite{Ashton:2018jfp} libraries can be used, alongside public samplers. In the main paper, we employed the {\sc{Dynesty}}~\cite{Speagle:2019ivv} nested sampler, as has been done in recent LVK analyses~\cite{LIGOScientific:2025yae}. We used 500 live points, which is less than that used for single-event Bayesian inference. Fewer live points were used since we are only interested in fitting one dimension: the astrophysical compactness of the detected CBCs. We also assumed that $p_{\mathrm{det}} = 1$ for all compactness values, which implies that we are implicitly assuming all existing GW detectors are equally sensitive to exotic binaries as BBHs.

Rather than assuming a specific functional form for the astrophysical distribution, $\Pi(\boldsymbol{\lambda}_{i} | \boldsymbol{\Lambda})$, we adopted a weakly modelled, data-driven approach, which makes minimal prior assumptions about the under-lying astrophysical population. We assumed a spline model with nodes placed equally within the astrophysical region of interest, and added some additional \emph{ghost} nodes, lying outside of the astrophysical region, in order to control edge effects. We used agnostic priors for the amplitude of each node by assuming a unit Gaussian distribution, as was done in e.g. Ref.~\cite{Golomb:2022bon}. We also checked the robustness of our population result by varying the number of nodes from 8-18, and consistent results were obtained.

%

\bibliography{refs.bib}

\end{document}